%% file: main.tex
\documentclass[10pt,twocolumn,letterpaper]{article}

\usepackage[final]{wacv} 

\usepackage{graphicx}
\usepackage{amsmath}
\usepackage{amssymb}
\usepackage{booktabs}
\usepackage{multirow}
\usepackage{array}
\usepackage{xcolor}
\usepackage[protrusion=false]{microtype}
\definecolor{wacvblue}{rgb}{0.21,0.49,0.74}
\usepackage[breaklinks,colorlinks,allcolors=wacvblue]{hyperref}

\definecolor{pend}{RGB}{140,140,140}



\def\wacvPaperID{1668}
\def\confName{WACV}
\def\confYear{2027}

\makeatletter
\renewcommand\section{\@startsection{section}{1}{\z@}{-8pt plus -2pt minus -1pt}{4pt plus 1pt minus 1pt}{\normalfont\large\bfseries}}
\renewcommand\subsection{\@startsection{subsection}{2}{\z@}{-6pt plus -2pt minus -1pt}{2pt plus 1pt}{\normalfont\normalsize\bfseries}}
\renewcommand\paragraph{\@startsection{paragraph}{4}{\z@}{3pt plus 1pt minus 1pt}{-1em}{\normalfont\normalsize\bfseries}}
\makeatother

\begin{document}

\title{CRS-Bench: A Reference-Relative Reliability Benchmark\\for Medical Image Encoders}
\author{
  Xingtao Lin$^{1}$\qquad
  Hangqi Ren$^{1}$\qquad
  Caiwan Sun$^{1}$\qquad
  You Chen$^{1,2}$\\[6pt]
  $^{1}$Vanderbilt University\qquad
  $^{2}$Vanderbilt University Medical Center (VUMC)
}
\maketitle

\begin{abstract}
Pretrained image encoders are central to medical image classification, where expert annotation is costly and task-specific cohorts are often limited. As the model space expands from general-purpose to broad-medical and specialty-specific encoders, selecting the representation becomes a substantive modeling decision. Clean-test discrimination alone is insufficient for this purpose: encoders with similar AUROC can differ in calibration, label efficiency, and stability under acquisition perturbations or distribution shift. We introduce \textbf{CRS-Bench}, a controlled benchmark for multi-objective medical encoder selection. CRS-Bench evaluates 15 pretrained encoder families across dermatology, ophthalmology, and radiology using ISIC 2019, APTOS 2019, and CheXpert, with CheXpert$\rightarrow$MIMIC-CXR as an observed institutional shift, yielding 17{,}575 controlled run records and 3{,}515 seed-aggregated metric rows. Each encoder is characterized along four operational reliability dimensions (discrimination, calibration, label efficiency, and robustness) and summarized by the \textbf{Clinical Reliability Score (CRS)}, a Pareto-aware, reference-relative score combining dominance, profile balance, and worst-axis performance. AUROC and CRS are positively associated but not decision-equivalent: \textbf{21 of 105 pairwise orderings reverse}, with a mean absolute rank displacement of 1.87. Paired-seed bootstrap analysis identifies PanDerm, MedSigLIP, and MedGemma as a stable leading reliability tier rather than a statistically resolved single leader. Reference-panel, scalarization, estimator, and cross-axis analyses further characterize criterion stability and information content. A matched 135-run adaptation arm keeps the leading three encoders intact ($\tau=0.562$, Top-3 overlap 1.000) while reordering the middle of the ranking, so frozen comparisons bound rather than determine post-adaptation behavior. CRS-Bench provides a controlled framework for selecting medical image encoders from their \textbf{multi-axis reliability profiles}, rather than from clean AUROC alone.
\end{abstract}

\section{Introduction}
\label{sec:intro}

Pretrained image encoders are a standard component of medical image classification, especially where expert annotation is expensive, task-specific cohorts are modest, and acquisition protocols vary across institutions. Large-scale pretraining transfers reusable visual structure into these low-data regimes and can reduce downstream supervision. The available model space has consequently become heterogeneous, spanning general-purpose encoders such as ResNet~\cite{he2016deep}, ViT~\cite{dosovitskiy2021image}, CLIP~\cite{radford2021learning}, DINOv2~\cite{oquab2023dinov2}, and MAE~\cite{he2022masked}; broad-medical models such as BiomedCLIP~\cite{zhang2023biomedclip}, MedSigLIP, and MedGemma~\cite{sellergren2025medgemma}; and specialty encoders for dermatology, ophthalmology, and radiology~\cite{yan2025panderm,zhou2023retfound,perezgarcia2025raddino,boecking2022making,tiu2022expert}. The resulting problem is how to select among substantially different encoders for a clinical task.

Current selection practice is dominated by clean-test discrimination, typically AUROC or AUPRC. Although necessary, discrimination is not sufficient to characterize an encoder for transfer. Representations with similar AUROC can differ materially in probability calibration, supervision demand, and stability under acquisition perturbations or distribution shift. An encoder may therefore rank highly on a clean split while remaining miscalibrated, supervision-intensive, or unstable under conditions that differ from the evaluation distribution~\cite{guo2017calibration,naeini2015obtaining,jiang2012calibrating,mehrtash2020confidence,castro2020causality,hendrycks2019benchmarking,ovadia2019trust}. Existing transfer and medical benchmarks measure subsets of these properties, but rarely all four under one matched clinical protocol, limiting guidance when selection targets a multidimensional reliability profile rather than a single predictive metric.

We introduce \textbf{CRS-Bench}, a controlled benchmark for multi-objective medical encoder selection. Fifteen pretrained encoder families are evaluated across dermatology, ophthalmology, and radiology using ISIC 2019, APTOS 2019, and CheXpert, with CheXpert$\rightarrow$MIMIC-CXR as an observed institutional shift. Matched downstream capacity, data splits, label fractions, perturbations, random seeds, and tuning budgets yield 17{,}575 run records and 3{,}515 seed-aggregated metric rows. Each encoder is characterized by four operational dimensions: \textbf{discrimination, calibration, label efficiency, and robustness}.

To summarize these trade-offs, we propose the \textbf{Clinical Reliability Score (CRS)}. CRS combines Pareto dominance, proximity to a balanced high-performing profile, and worst-axis performance. Because the score is comparative, its reference panel and normalization anchors are explicit and fixed, allowing later encoders to be evaluated without changing previously reported values. We then examine whether this criterion remains stable under reference-panel and scalarization perturbations, whether its constituent dimensions contribute information beyond AUROC, how sampling uncertainty affects the leading set, and how far the ordering persists under downstream adaptation and excluded data.

The results establish that clean discrimination and multi-axis reliability are related but not decision-equivalent: \textbf{21 of 105 pairwise encoder orderings reverse} between AUROC and CRS, with mean absolute displacement 1.87. Bootstrap analysis supports a stable leading tier of PanDerm, MedSigLIP, and MedGemma rather than a statistically resolved single leader. The benchmark further reveals systematic differences in encoder behavior: specialty pretraining can yield large matched-domain gains with heterogeneous transfer; calibration is often substantially improved post hoc without changing AUROC; intermediate representations can outperform final-layer features; corruption can preserve discrimination while degrading confidence; and medical pretraining provides its largest relative advantage when labels are scarce.

\paragraph{Contributions.}
\textbf{(i) CRS-Bench.} A controlled multi-domain evaluation of 15 pretrained encoders across four complementary reliability dimensions.
\textbf{(ii) CRS.} A Pareto-aware, reference-relative summary whose published values remain comparable as new encoders are evaluated.
\textbf{(iii) Criterion validation.} Systematic analysis of reference stability, construct informativeness, scalarization sensitivity, sampling uncertainty, estimator dependence, and decision consequences beyond AUROC.
\textbf{(iv) Encoder analysis.} Empirical characterization of specialization, calibration, feature depth, label scarcity, perturbation robustness, prompting, and downstream adaptation.

\section{Related Work}
\label{sec:related}

\noindent\textbf{Pretrained encoders for medical imaging.} Generalist models use
supervised natural-image pretraining (ResNet~\cite{he2016deep},
ViT~\cite{dosovitskiy2021image}), image--text contrastive learning
(CLIP~\cite{radford2021learning}, SigLIP~\cite{zhai2023siglip}), or
self-supervision (DINOv2~\cite{oquab2023dinov2}, MAE~\cite{he2022masked}).
Broad-medical encoders adapt these recipes to biomedical data
(BiomedCLIP~\cite{zhang2023biomedclip}, LLaVA-Med~\cite{li2023llava}, MedSigLIP,
MedGemma~\cite{sellergren2025medgemma}), while RETFound~\cite{zhou2023retfound},
PanDerm~\cite{yan2025panderm}, RAD-DINO~\cite{perezgarcia2025raddino},
BioViL~\cite{boecking2022making}, and CheXzero~\cite{tiu2022expert} target
particular clinical domains. Prior comparisons generally cover narrower model sets and emphasize clean discrimination or task-specific adaptation.

\noindent\textbf{Benchmarking reliability.} VTAB~\cite{zhai2019vtab} and
ELEVATER~\cite{li2022elevater} measure representation transfer, ImageNet-C
standardizes corruption robustness~\cite{hendrycks2019benchmarking}, and medical
benchmarks emphasize few-shot adaptation, modality breadth, radiology, or
fairness~\cite{wang2023medfmc,mo2024medvtab,wu2025radfm,jin2024fairmedfm}.
Calibration and uncertainty studies further show that high discrimination does
not guarantee reliable probabilities and that confidence degrades under
shift~\cite{guo2017calibration,ovadia2019trust,jiang2012calibrating,mehrtash2020confidence,castro2020causality}. CRS-Bench instead measures four reliability dimensions under one controlled clinical protocol and defines its aggregate against an explicit reference panel.

\begin{figure}[t]
\centering
\includegraphics[width=\linewidth]{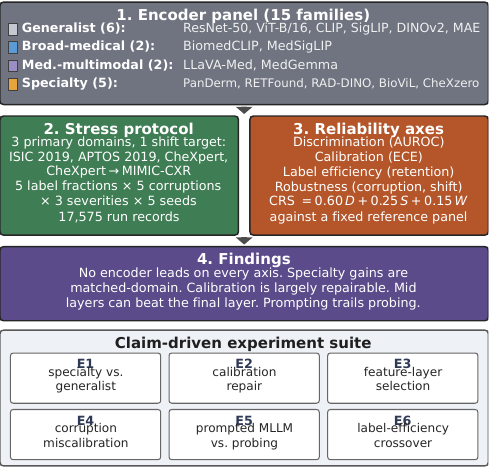}
\caption{\textbf{CRS-Bench evaluation pipeline.} Fifteen encoder families are evaluated under one controlled protocol, and four reliability dimensions define the profile summarized by CRS.}
\label{fig:pipeline}
\end{figure}

\section{The CRS-Bench Benchmark}
\label{sec:bench}

CRS-Bench separates three methodological tasks: \emph{measurement}, \emph{score
validation}, and \emph{behavioral interpretation} (Fig.~\ref{fig:pipeline}).
Secs.~\ref{sec:panel}--\ref{sec:stress} measure four reliability dimensions under
matched frozen probing, label scarcity, acquisition perturbations, and
institutional shift. Secs.~\ref{sec:crs}--\ref{sec:suite} define the
reference-relative CRS and test reference stability (panel perturbation and
insertion), construct informativeness/scalarization (correlations and weight
perturbations), sampling uncertainty (paired-seed bootstrap), estimator
dependence, and ranking transfer (LoRA and held-out data). Sec.~\ref{sec:mechanisms}
defines E1--E6 for specialization, calibration, feature depth, corruption,
prompting, and label scarcity. This structure separates what is measured, how
the aggregate is validated, and what the benchmark reveals about encoders.

\subsection{Selection Problem and Encoder Panel}
\label{sec:panel}

CRS-Bench studies \emph{controlled encoder selection}: given a declared clinical
evaluation suite, how do pretrained representations compare when selection is
based jointly on several observable reliability properties under matched
downstream conditions? For
encoder panel $\mathcal{M}=\{m_1,\ldots,m_K\}$, representation
$z=m_k(x)\in\mathbb{R}^{d_k}$, and lightweight probe $\hat y=h_\theta(z)$, we
measure
\begin{equation}
\mathbf{s}^{(k)}=(s_{\mathrm{disc}}^{(k)},s_{\mathrm{cal}}^{(k)},
 s_{\mathrm{LE}}^{(k)},s_{\mathrm{rob}}^{(k)}),
\label{eq:profile}
\end{equation}
for discrimination, calibration, label efficiency, and robustness. CRS
(\cref{sec:crs}) is a reference-relative summary of this measured profile rather
than an estimator of universal clinical utility. Candidate models may evolve
while the comparison reference remains explicit, so later evaluations do not
implicitly redefine earlier scores.

The 15-family panel spans general-purpose, broad-medical or medical-multimodal,
and specialty-specific encoders (\cref{tab:models}). MedSigLIP and MedGemma are
kept distinct because we probe different released visual representations: the
standalone MedSigLIP tower at native 448 resolution and the MedGemma-4B vision
stack after projection into its generative interface. The frozen panel contains
43/45 encoder--dataset cells; BioViL and CheXzero have no APTOS cell because
their released interfaces are chest-radiograph specific. Their aggregate values
use the two observed domains and are never imputed.

\begin{table}[t]
\centering
\scriptsize
\renewcommand{\arraystretch}{0.72}
\caption{The 15 encoder families by pretraining scope. ``Input'' is the released input size used by each encoder (\cref{sec:protocol}). Domain marks the matched specialist domain: Derm.\ = ISIC, Eye = APTOS, Rad.\ = CheXpert/MIMIC-CXR. $^\dagger$ indicates no APTOS frozen cell.}
\label{tab:models}
\setlength{\tabcolsep}{2.2pt}
\begin{tabular}{@{}lllcl@{}}
\toprule
Model & Pretraining data & Paradigm & Input & Domain \\
\midrule
\multicolumn{5}{@{}l}{\emph{Generalist natural-image encoders}}\\
ResNet-50~\cite{he2016deep}        & ImageNet-1K   & Supervised  & 224 & n/a \\
ViT-B/16~\cite{dosovitskiy2021image} & ImageNet-21K & Supervised & 224 & n/a \\
CLIP ViT-B~\cite{radford2021learning} & WIT-400M   & Contrastive & 224 & n/a \\
SigLIP ViT-B~\cite{zhai2023siglip} & WebLI         & Contrastive & 224 & n/a \\
DINOv2 ViT-B~\cite{oquab2023dinov2} & LVD-142M     & Self-sup.\  & 224 & n/a \\
MAE ViT-B~\cite{he2022masked}      & ImageNet-1K   & Self-sup.\  & 224 & n/a \\
\midrule
\multicolumn{5}{@{}l}{\emph{Broad-medical and medical-multimodal encoders}}\\
BiomedCLIP~\cite{zhang2023biomedclip} & PMC-15M    & Contrastive & 224 & Med.\ \\
LLaVA-Med~\cite{li2023llava}       & PMC, instruct.\ & MLLM     & 336 & Med.\ \\
MedSigLIP~\cite{sellergren2025medgemma} & Med.\ image/text & Contrastive & 448 & Med.\ \\
MedGemma~\cite{sellergren2025medgemma}  & Med.\ image/text, IT & MLLM & own & Med.\ \\
\midrule
\multicolumn{5}{@{}l}{\emph{Specialty-specific encoders}}\\
PanDerm~\cite{yan2025panderm}      & Skin images & Self-sup.\  & 224 & Derm.\ \\
RETFound~\cite{zhou2023retfound}   & Fundus / OCT  & MAE-SSL     & 224 & Eye \\
RAD-DINO~\cite{perezgarcia2025raddino} & Chest X-rays & DINO-SSL & 224 & Rad.\ \\
BioViL$^\dagger$~\cite{boecking2022making} & MIMIC-CXR & Contrastive & 480 & Rad.\ \\
CheXzero$^\dagger$~\cite{tiu2022expert} & MIMIC-CXR & Contrastive & 320 & Rad.\ \\
\bottomrule
\end{tabular}
\end{table}

\subsection{Controlled Representation Evaluation}
\label{sec:protocol}

Frozen probing defines the primary estimand: representation quality under a
common downstream decision rule. Fixing probe family and optimization budget
reduces confounding from task-specific optimization and isolates differences in
the pretrained feature space. LoRA is evaluated as a separate regime
(\cref{sec:adapt}) that measures how much of the frozen ordering persists after
task-specific adaptation.

Each encoder uses the input transform required by its released interface
(\cref{tab:models}). After feature extraction, data splits, label subsamples,
perturbation seeds, probe capacity, hyperparameter budget, wall-clock cap, and
five seeds $s\in\{0,\ldots,4\}$ are matched. Features are standardized with
training-split statistics. The primary probe is $\ell_2$-regularized logistic
regression ($\lambda=1/N_{\mathrm{train}}$, L-BFGS), one-vs-rest for CheXpert;
a two-layer MLP and $k{=}5$ nearest neighbor provide probe-dependence checks.
For specialists that accept the common $224$-pixel ImageNet transform, a
standardized-input sensitivity changes AUROC by $<0.006$ (supplement).

\noindent\textbf{Two-tier evaluation.} A factorial core varies five label
fractions, five perturbation types, three severity levels, and five shared seeds
on representative encoders for ISIC and APTOS. A broad sweep evaluates all 15
families on ISIC, APTOS, and CheXpert, with CheXpert probes additionally tested
on MIMIC-CXR without retraining. The benchmark contains 17{,}575 runs and 3{,}515 seed-aggregated metric rows.

\noindent\textbf{Adaptation regime.} LoRA uses rank $8$, $\alpha=16$, dropout
$0.05$, learning rate $2\times10^{-4}$, eight epochs, attention/MLP projection
adapters, and three seeds. Each encoder--dataset--seed triple is evaluated over
the same label-fraction and perturbation grid (135 encoder--dataset--seed runs), with all four
dimensions recomputed after adaptation. The analysis therefore tests transfer
of the frozen selection ordering rather than redefining CRS around fine-tuning.

\subsection{Reliability Dimensions}
\label{sec:axes}

Each dimension is computed per dataset before aggregation in \cref{sec:crs}.
They are operationally distinct but not assumed statistically independent;
empirical dependence is tested in \cref{sec:axes-results}.

\noindent\textbf{Discrimination.} We use macro-AUROC,
\begin{equation}
\mathrm{AUROC}_{\mathrm{macro}}=\frac{1}{C}\sum_{c=1}^{C}\mathrm{AUROC}_c,
\end{equation}
computed one-vs-rest for multiclass tasks and per-label then macro-averaged for
CheXpert. Macro AUPRC and sensitivity at fixed specificity are secondary
metrics. Because APTOS grades are ordinal, QWK and ordinal MAE are additionally
substituted for the discrimination axis in \cref{sec:estimators}.

\noindent\textbf{Calibration.} ECE with $B{=}15$ fixed-width bins is
\begin{equation}
\mathrm{ECE}=\sum_{b=1}^{B}\frac{|S_b|}{N}
\bigl|\mathrm{acc}(S_b)-\mathrm{conf}(S_b)\bigr| .
\label{eq:ece}
\end{equation}
Confidence is maximum class probability for multiclass tasks; CheXpert ECE is
macro-averaged across labels. Fixed-width ECE provides a deterministic common estimator across the model
panel. Adaptive ECE tests sensitivity to binning, while Brier score evaluates
whether the conclusions persist under a proper scoring rule
(\cref{sec:estimators}).

\noindent\textbf{Label efficiency.} Absolute low-label AUROC mixes sample
efficiency with attainable discrimination. We instead measure retention of each
encoder's own full-label performance. For
$\mathcal{P}=\{0.01,0.05,0.10,0.25\}$,
\begin{equation}
\mathrm{LE}=\frac{1}{|\mathcal{P}|}\sum_{\rho\in\mathcal{P}}
\frac{A(\rho)}{A(1)} ,
\label{eq:le}
\end{equation}
where $A(\rho)$ and $A(1)$ are macro-AUROC at fraction $\rho$ and full labels
for the same encoder--dataset pair. The full-label point appears only in the
denominator, so LE measures preservation under label scarcity rather than
absolute attainment. This distinction matters because conventional area under
the label-fraction curve (AULC) rewards a high full-label endpoint even when a
model loses a similar fraction of performance as labels are removed. We compare
AULC and relative retention directly in \cref{sec:axes-results} rather than
assuming the two encode distinct information. Ratios slightly above one are kept
(maximum $1.003$) and bounded only by the fixed CRS normalization; a log-spaced
trapezoidal variant is reported in the supplement.

\noindent\textbf{Robustness.} With clean AUROC $A_0$ and perturbed AUROC $A_j$,
let $\delta_j=\max(0,(A_0-A_j)/A_0)$ and let $\bar\delta,\delta_{\max}$ be the
mean and worst perturbation drops. CheXpert additionally uses
$\delta_{\mathrm{shift}}$ from MIMIC-CXR. We define
\begin{equation}
R=\max\!\left(0,1-\frac{\bar\delta+\delta_{\max}+
\mathbf{1}_{\mathrm{shift}}\delta_{\mathrm{shift}}}{2+\mathbf{1}_{\mathrm{shift}}}\right),
\label{eq:robust}
\end{equation}
with $\mathbf{1}_{\mathrm{shift}}=1$ for CheXpert and $0$ otherwise. The
denominator counts the perturbation statistics that were actually observed, so a
cell without a natural shift is scored as the mean of its two corruption terms
rather than being charged a third term that was never measured. Robustness is
consequently comparable across cells in construction while remaining
heterogeneous in evidence: only CheXpert contributes an observed institutional
shift. We therefore report a matched corruption-only variant, in which every cell
is scored from corruption alone, as the direct test of whether that heterogeneity
carries the ranking (\cref{sec:estimators}); E4 separately quantifies calibration
degradation under corruption.

\subsection{Stress Conditions}
\label{sec:stress}

\noindent\textbf{Acquisition-quality perturbations.} Five perturbations are
applied at three severities: Gaussian blur, JPEG compression, contrast
reduction, Rician noise~\cite{gudbjartsson1995rician}, and field-of-view crop.
Rician noise uses $\tilde{x}=\sqrt{(x+n_r)^2+n_i^2}$,
$n_r,n_i\sim\mathcal{N}(0,\sigma^2)$. Perturbation seeds are derived from a
SHA-256 hash of the image identifier, ensuring that the same transformed image
reaches every encoder. These stresses model acquisition
and image-quality degradation, not changes in anatomy, pathology, or population.

\noindent\textbf{Observed institutional shift.} CheXpert-trained probes are
applied to MIMIC-CXR without retraining. The datasets share 14 labels but differ
in institution, population, protocol, and labeling, testing preservation of
discriminative structure across institutions.

\subsection{Reference-Relative Clinical Reliability Score}
\label{sec:crs}

After orienting all four dimensions so larger is better, CRS combines Pareto
dominance over a declared reference panel, proximity to an ideal profile, and
worst-axis performance,
\begin{align}
D^{(k)}&=\frac{1}{n_k}\!\sum_{j\in\mathcal{R}\setminus\{k\}}
\mathbf{1}[\tilde{\mathbf{s}}^{(k)}\succ\tilde{\mathbf{s}}^{(j)}],
\label{eq:dom}\\
S^{(k)}&=1-\sum_a w_a(1-\tilde{s}^{(k)}_a)^2,
\quad W^{(k)}=\min_a\tilde{s}^{(k)}_a,
\label{eq:sw}\\
\mathrm{CRS}_{\mathcal{R},\mathcal{A}}^{(k)}
&=\alpha_DD^{(k)}+\alpha_SS^{(k)}+\alpha_WW^{(k)}.
\label{eq:crs}
\end{align}
Here $D$ measures Pareto-consistent improvement over the reference panel, $S$
measures proximity to the ideal profile, and $W$ retains sensitivity to the
weakest axis. The three terms provide complementary summaries without treating
the four raw metrics as directly commensurate.
$n_k=|\mathcal{R}|-\mathbf{1}[k\in\mathcal{R}]$.

\noindent\textbf{Reference standard.} Candidate-dependent min--max scaling and
dominance make an unchanged encoder's score move when new candidates are
inserted. CRS therefore fixes a reference
$(\mathcal{R},\mathcal{A})$, where $\mathcal{R}$ supplies dominance peers and
$\mathcal{A}=\{(l_a,u_a)\}_a$ the normalization anchors,
\begin{equation}
\tilde{s}_a=\mathrm{clip}\!\left(\frac{s_a-l_a}{u_a-l_a},0,1\right).
\label{eq:clip}
\end{equation}
The 15-family panel under \cref{sec:protocol} defines the aggregate reference.
A future encoder is normalized by the same anchors and compared with the same
peers, so previously reported scores do not change. Per-domain CRS uses an
analogous domain-specific reference and is interpreted only within that domain.
Repeated anchor saturation signals when a successor reference is needed.

\noindent\textbf{Scalarization preferences.} Axis weights $w_a$ affect only the
ideal-point term and are uniform ($1/4$) by default. Component weights
$(\alpha_D,\alpha_S,\alpha_W)=(0.60,0.25,0.15)$ prioritize Pareto-consistent
comparative evidence while retaining ideal-profile and worst-axis penalties.
Both are explicit design parameters rather than learned clinical utilities.
The two weight families are perturbed separately in \cref{sec:axes-results};
our analysis concerns the stability of the induced ordering, not optimality of a
particular coefficient vector.

\subsection{Validation Design}
\label{sec:suite}

We evaluate five properties of the selection criterion. \textbf{Reference
stability/extensibility} uses 15 leave-one-encoder-out panels, 200 random subsets
at sizes 12 and 10, and 30 random 12+3 insertion trials comparing cohort
recomputation with the fixed reference (\cref{sec:extensible}). \textbf{Construct
informativeness} uses Spearman correlations at encoder and encoder--dataset
levels and directly contrasts relative-retention LE with AULC; \textbf{weight
sensitivity} separately perturbs 500 axis-weight and 500 component-weight draws
(\cref{sec:axes-results}). \textbf{Decision consequence} compares AUROC and CRS
using Kendall $\tau$, Spearman $\rho$, pairwise reversals, rank displacement, and
Top-$k$ overlap (\cref{sec:divergence}).

\textbf{Sampling uncertainty} uses 10{,}000 paired-seed bootstrap replicates over
the five shared seeds and 43 observed cells, recomputing anchors and CRS within
each replicate (\cref{sec:leaderboard}). Estimator substitutions use adaptive
ECE/Brier, APTOS QWK/ordinal MAE, and matched corruption-only robustness
(\cref{sec:estimators}). LoRA and leave-one-dataset-out/MIMIC analyses test
ranking transfer under adaptation and excluded data without redefining
future-task prediction as the CRS objective (\cref{sec:adapt,sec:external}).

\subsection{Encoder-Level Analysis Suite}
\label{sec:mechanisms}

Six analyses explain the profiles themselves. \textbf{E1} separates specialist
matched-domain gain from transfer using
$\Delta_{\mathrm{spec}}=\mathrm{AUROC}_{\mathrm{in}}-\mathrm{AUROC}_{\mathrm{out}}$
and $\Delta_{\mathrm{ood}}=\mathrm{CRS}_{\mathrm{in}}-\mathrm{CRS}_{\mathrm{out}}$.
\textbf{E2} fits temperature, vector, isotonic, and Dirichlet calibration on a
held-out calibration split and recomputes ECE/Brier while checking unchanged
AUROC. \textbf{E3} probes transformer features at 25\%, 50\%, 75\%, and 100\%
depth. \textbf{E4} pairs AUROC drop with ECE, confidence shift, and
overconfidence-under-shift. \textbf{E5} compares frozen probing with
simple, multiple-choice, chain-of-thought, refined, and few-shot prompts for
LLaVA-Med/MedGemma, reporting parse coverage. \textbf{E6} reports AUROC at
1\%/10\% labels and the first label fraction matching ViT-B/16. Together, these analyses characterize the empirical mechanisms underlying the observed
profiles rather than treating scalar rank as the sole endpoint.

\section{Experimental Setup}
\label{sec:setup}

\noindent\textbf{Datasets.} Three primary datasets and one external shift
target: CheXpert~\cite{irvin2019chexpert} (191{,}229 chest radiographs, 14
labels, multilabel); ISIC 2019~\cite{codella2019isic,hernandez2024bcn20000}
(25{,}331 dermoscopy images, 8 classes); APTOS 2019~\cite{aptos2019} (3{,}662
fundus photographs, 5 ordinal grades); and
MIMIC-CXR~\cite{johnson2019mimiccxr} as the institutional-shift target for
CheXpert-trained probes. Together they span distinct modalities, anatomical
scales, and acquisition pipelines.

\noindent\textbf{Implementation.} Following \cref{sec:protocol}, all encoders
are frozen and evaluated in a single feature pass through their required input
transform, with shared splits, seeds, probe capacity, and tuning budget. For generative VLMs in E5 we use
simple, multiple-choice, chain-of-thought, refined, and few-shot prompts,
parsed deterministically with coverage recorded per condition. Frozen
experiments run on a single GB10 DGX Spark; the 135 adaptation runs are
distributed over a compute cluster.

\section{Results}
\label{sec:results}

Our evaluation addresses two goals. For criterion validity,
Sec.~\ref{sec:leaderboard} combines four-axis profiles with 10{,}000 bootstrap
replicates; Sec.~\ref{sec:extensible} tests reference stability using 415 panel
perturbations and 30 insertion trials; Sec.~\ref{sec:axes-results} tests
informativeness and scalarization using cross-axis correlations and 1{,}000
weight perturbations; Sec.~\ref{sec:divergence} quantifies AUROC--CRS decision
differences; and Secs.~\ref{sec:adapt}--\ref{sec:estimators} test LoRA transfer,
held-out/MIMIC-CXR transfer, and estimator dependence. For encoder behavior,
Secs.~\ref{sec:domain-results}--\ref{sec:prompt-results} execute E1 plus
institutional shift, E2/E4, E3/E6, and E5 to study specialization/transfer,
calibration/corruption, feature depth/label scarcity, and prompting/probing.

\begin{figure}[!b]
\centering
\includegraphics[width=0.72\linewidth]{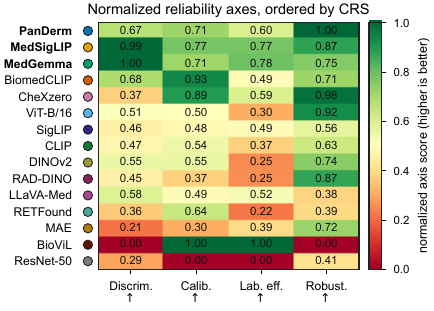}
\caption{\textbf{Normalized reliability profiles.} Encoders are ordered by CRS; higher is better on every axis. Similar discrimination can coexist with different calibration, label retention, and robustness.}
\label{fig:axes}
\end{figure}

\subsection{Multi-Axis Profiles and Sampling Uncertainty}
\label{sec:leaderboard}

Clean AUROC and CRS yield related but distinct orderings
(\cref{tab:leaderboard,fig:axes}). MedGemma/MedSigLIP define the AUROC frontier
(0.909/0.908), whereas PanDerm and MedSigLIP are separated by 0.002 in CRS
(0.703/0.701).
Figure~\ref{fig:axes} explains the reordering: PanDerm pairs competitive
discrimination with the strongest robustness; MedSigLIP/MedGemma concentrate
more advantage in discrimination and low-label retention; BiomedCLIP is
comparatively strong in calibration; and CheXzero combines lower discrimination
with favorable calibration and robustness. Per-domain leaders also change, with PanDerm leading on ISIC CRS, MedGemma on APTOS, and MedSigLIP on CheXpert, so the aggregate ranking summarizes distinct reliability profiles rather than a rescaled accuracy statistic.

Across 10{,}000 paired-seed bootstrap replicates, PanDerm/MedSigLIP attain rank 1
with probability 0.664/0.336 and their paired CRS difference has 95\% interval
$[-0.097,0.017]$. PanDerm, MedSigLIP, and MedGemma each have
$P(\mathrm{top\mbox{-}3})=1.000$, whereas every other encoder has probability
zero. The bootstrap therefore supports a three-model leading tier rather than a
statistically resolved single leader. Because anchors are recomputed within each replicate,
the reported intervals describe the sampling distribution of the full scoring
procedure rather than intervals centered on the fixed-anchor point estimate.

\begin{table*}[t]
\centering
\scriptsize
\renewcommand{\arraystretch}{0.60}
\setlength{\tabcolsep}{1.45pt}
\caption{Encoder reliability by dataset. CRS uses domain-specific references per dataset and the global reference in aggregate. BioViL/CheXzero lack APTOS. CI is the paired-seed sampling interval; $P_1$ is rank-1 probability.}
\label{tab:leaderboard}
\begin{tabular}{@{}ll cc cc cc cc c c@{}}
\toprule
& & \multicolumn{2}{c}{ISIC} & \multicolumn{2}{c}{APTOS} & \multicolumn{2}{c}{CheXpert} & \multicolumn{2}{c}{Aggregate} & & \\
\cmidrule(lr){3-4}\cmidrule(lr){5-6}\cmidrule(lr){7-8}\cmidrule(lr){9-10}
Model & Grp & AUROC & CRS & AUROC & CRS & AUROC & CRS & AUROC & CRS & CI & $P_1$ \\
\midrule
PanDerm & S & 0.957 & \textbf{0.583} & 0.922 & 0.685 & 0.733 & 0.346 & 0.871 & \textbf{0.703} & [0.600,0.720] & \textbf{0.66} \\
MedSigLIP & BM & 0.957 & 0.485 & 0.951 & 0.755 & \textbf{0.816} & \textbf{0.524} & 0.908 & 0.701 & [0.682,0.703] & 0.34 \\
MedGemma & MM & \textbf{0.960} & 0.333 & \textbf{0.958} & \textbf{0.763} & 0.808 & 0.359 & \textbf{0.909} & 0.645 & [0.598,0.647] & 0.00 \\
BiomedCLIP & BM & 0.899 & 0.449 & 0.908 & 0.454 & 0.812 & 0.246 & 0.873 & 0.424 & [0.406,0.522] & 0.00 \\
CheXzero & S & 0.886 & 0.253 & n/a & n/a & 0.788 & 0.418 & 0.837 & 0.398 & [0.391,0.403] & 0.00 \\
ViT-B/16 & G & 0.926 & 0.323 & 0.906 & 0.297 & 0.727 & 0.332 & 0.853 & 0.320 & [0.228,0.443] & 0.00 \\
SigLIP & G & 0.935 & 0.409 & 0.920 & 0.220 & 0.686 & 0.142 & 0.847 & 0.298 & [0.278,0.305] & 0.00 \\
CLIP & G & 0.927 & 0.396 & 0.898 & 0.303 & 0.720 & 0.154 & 0.848 & 0.284 & [0.248,0.352] & 0.00 \\
DINOv2 & G & 0.935 & 0.480 & 0.912 & 0.363 & 0.725 & 0.202 & 0.857 & 0.266 & [0.168,0.352] & 0.00 \\
RAD-DINO & S & 0.891 & 0.165 & 0.846 & 0.138 & 0.803 & 0.244 & 0.847 & 0.251 & [0.142,0.331] & 0.00 \\
LLaVA-Med & MM & 0.938 & 0.364 & 0.924 & 0.250 & 0.723 & 0.078 & 0.861 & 0.241 & [0.231,0.247] & 0.00 \\
RETFound & S & 0.900 & 0.257 & 0.907 & 0.098 & 0.702 & 0.186 & 0.836 & 0.187 & [0.131,0.225] & 0.00 \\
MAE & G & 0.896 & 0.279 & 0.875 & 0.137 & 0.687 & 0.145 & 0.819 & 0.184 & [0.172,0.194] & 0.00 \\
BioViL & S & 0.809 & 0.125 & n/a & n/a & 0.782 & 0.323 & 0.795 & 0.125 & [0.125,0.125] & 0.00 \\
ResNet-50 & G & 0.881 & 0.144 & 0.880 & 0.160 & 0.724 & 0.097 & 0.828 & 0.072 & [0.071,0.085] & 0.00 \\
\bottomrule
\end{tabular}
\end{table*}

\subsection{Reference Stability and Incremental Extensibility}
\label{sec:extensible}
Panel perturbation tests dependence on current peers; incremental insertion tests
whether future candidates move published scores. Across 415 perturbations,
Kendall $\tau$ is 0.975/0.926/0.891 for leave-one-out, size-12, and size-10
panels, with Top-3 overlap 1.000/0.985/0.958. Agreement weakens gradually as the
panel is thinned, but the leading tier remains stable. In 30 random 12+3 trials
(90 insertions), cohort renormalization moves existing scores by
mean/median/maximum 0.053/0.042/0.167, whereas the fixed-reference rule yields
exactly zero displacement. The zero displacement under fixed-reference insertion is a definitional
invariance rather than an empirical coincidence: a future encoder requires only
its own evaluation and cannot alter previously published scores.

\subsection{Construct Informativeness and Scalarization Stability}
\label{sec:axes-results}
\noindent\textbf{Axis informativeness.} AULC label efficiency is nearly a
re-expression of full-label AUROC ($\rho=0.986$ across encoders; 0.977 across 43
cells). Relative retention (\cref{eq:le}) reduces these to 0.339 and $-0.342$,
consistent with the intended interpretation of LE as relative performance retention under reduced supervision. The remaining
dimensions are dependent but non-interchangeable: the largest $|\rho|$ is
0.625/0.460 at encoder/cell level, while discrimination versus oriented
calibration changes from 0.764 on APTOS to $-0.279$ on CheXpert and $-0.029$ on
ISIC.

\noindent\textbf{Scalarization stability.} Across 500 axis-weight and 500
component-weight draws, mean Kendall agreement is 0.986/0.985 (minimum
0.943/0.924) and Top-3 is unchanged in all 1{,}000 draws; alternative axis
weights $(0.30,0.30,0.15,0.25)$ give $\tau=1.000$. These perturbations support stability of the leading set, but do not imply
uniqueness or optimality of the default coefficient vector.

\subsection{Decision Consequences beyond AUROC}
\label{sec:divergence}
\begin{figure*}[t]
\centering
\includegraphics[width=0.80\textwidth]{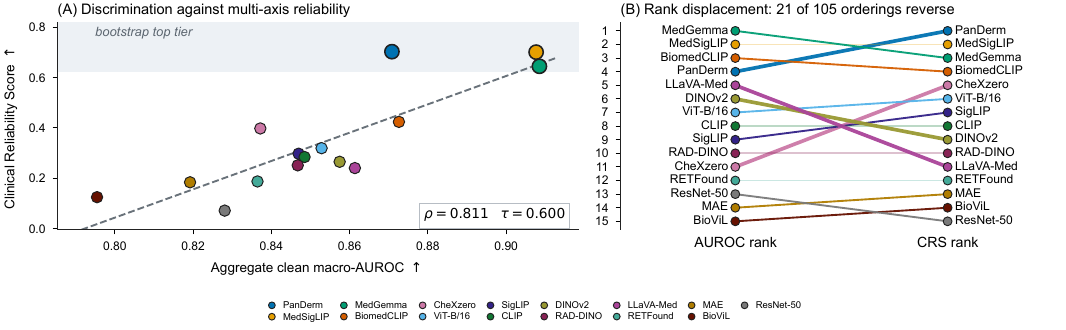}
\caption{\textbf{AUROC versus multi-axis reliability.} (A) Aggregate AUROC versus CRS ($\rho=0.811$, $\tau=0.600$). (B) Rank displacement: 21/105 pairwise orderings reverse (mean 1.87). The complete key identifies every model; panel (b) also labels each model directly.}
\label{fig:divergence}
\end{figure*}

Because discrimination is one CRS dimension, positive association is expected;
the question is whether the remaining axes change selections. Across 15
encoders, $\tau=0.600$, $\rho=0.811$, 21/105 pairwise choices reverse, mean
absolute displacement is 1.87 ranks, and Top-3/Top-5 overlap is 0.667/0.800
(\cref{fig:divergence}). CheXzero moves 11$\rightarrow$5, LLaVA-Med
5$\rightarrow$11, PanDerm 4$\rightarrow$1, and DINOv2 6$\rightarrow$9. CRS is
therefore correlated with, but not decision-equivalent to, clean discrimination;
the disagreement is distributed across the ordering rather than being confined
to a single change at the top.

\subsection{Transfer of the Ranking under Adaptation}
\label{sec:adapt}
Frozen probing fixes the downstream decision rule and therefore isolates the
pretrained representation; LoRA changes the representation itself. Agreement
between the two regimes consequently tests persistence of the frozen ordering
after task-specific optimization rather than serving as another estimate of the
same quantity. The matched adaptation arm recomputes discrimination,
calibration, label efficiency, robustness, and CRS on the same evaluation grid.
Adaptation raises clean discrimination in every domain, by
$+0.026/{+}0.030/{+}0.010$ mean macro-AUROC on ISIC/APTOS/CheXpert, and preserves
the discrimination ordering unevenly: Kendall $\tau=0.829/0.667/0.448$ with
Top-3 overlap $0.667$ in each domain, and the matched-domain leader changes in
all three (MedGemma to MedSigLIP on ISIC and APTOS, MedSigLIP to RAD-DINO on
CheXpert). The four-axis comparison is more demanding: against the frozen CRS
the adapted ranking gives $\tau=0.562$ and $\rho=0.707$ with a mean absolute
rank shift of 2.0. PanDerm is retained at rank 1 and the leading three are
unchanged (Top-3 overlap 1.000), whereas the middle reorganizes, with
BiomedCLIP falling from 4 to 13 and LLaVA-Med rising from 11 to 4. Frozen
rankings therefore characterize the pretrained representation and bound, rather
than determine, the ordering after parameter-efficient adaptation.

\subsection{External Validity of a Reference-Relative Score}
\label{sec:external}
Leave-one-dataset-out CRS correlates with the held-out ranking at
$\rho=0.550/0.709/0.271$ for ISIC/APTOS/CheXpert, with Top-3 overlap
1.000/0.333/0.667 and the held-out winner recovered in no fold. Removing MIMIC-CXR from CRS and
predicting MIMIC AUROC gives $\rho=0.529$, versus 0.864 for CheXpert AUROC. The
latter is more directly aligned because the target is discrimination on a
closely related radiographic dataset. These tests therefore bound CRS as a
multi-objective selection summary under a declared suite rather than a universal
predictor of unseen-task AUROC.

\subsection{Sensitivity to Metric Representation}
\label{sec:estimators}
Adaptive ECE leaves the ranking unchanged ($\tau=1.000$) and Brier gives
$\tau=0.981$, preserving the leading tier; the primary conclusion is therefore not determined by the fixed-width binning scheme. On APTOS, QWK/ordinal-MAE give $\tau=0.923/0.949$ with Top-3
preserved, indicating that the leading set is not an artifact of representing the ordinal task with macro-AUROC. Matched
corruption-only robustness gives $\tau=0.962$ with Top-3/Top-5 preserved, although
absolute CRS moves more (mean 0.031; maximum 0.171). Thus the two robustness
variants are different measurements that support the same leading set.

\subsection{Domain Specialization and Cross-Domain Transfer}
\label{sec:domain-results}
\label{sec:domain-findings}
\begin{figure}[tbp]
\centering
\begin{subfigure}[t]{0.485\linewidth}
  \centering
  \includegraphics[width=\linewidth]{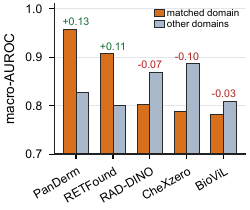}
  \caption{Matched vs. other-domain AUROC.}
\end{subfigure}\hfill
\begin{subfigure}[t]{0.485\linewidth}
  \centering
  \includegraphics[width=\linewidth]{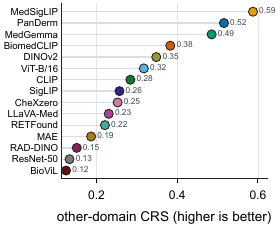}
  \caption{Other-domain CRS by encoder.}
\end{subfigure}
\caption{\textbf{Specialization versus transfer (E1).} Matched-domain discrimination and cross-domain reliability are complementary. Panel (b) ranks and labels every encoder; values are printed beside the corresponding model markers.}
\label{fig:specialization}
\end{figure}

Specialty pretraining yields substantial but heterogeneous matched-domain gains
(\cref{fig:specialization}). PanDerm is the clearest case: ISIC AUROC is 0.957
versus 0.828 out of domain ($+0.130$), with only a 0.068 out-of-domain CRS
penalty. RETFound gains ophthalmic discrimination without a commensurate
balanced-reliability advantage. For radiology specialists, raw in-minus-out
AUROC is less comparable because CheXpert is multilabel; the institutional-shift analysis provides a more directly interpretable transfer setting. Within this panel, broad-medical encoders are more
consistent across domains, while PanDerm shows that specialization need not
imply narrow transfer.

\begin{figure}[tbp]
\centering
\begin{subfigure}[t]{0.485\linewidth}
  \centering
  \includegraphics[width=\linewidth]{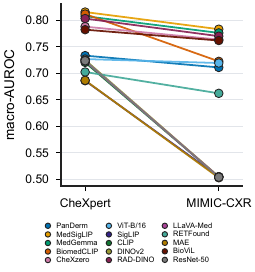}
  \caption{Source vs. shifted AUROC.}
\end{subfigure}\hfill
\begin{subfigure}[t]{0.485\linewidth}
  \centering
  \includegraphics[width=\linewidth]{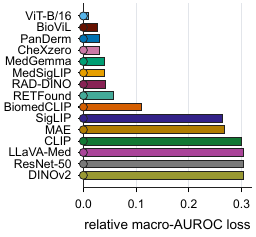}
  \caption{Relative AUROC loss.}
\end{subfigure}
\caption{\textbf{Institutional shift.} CheXpert$\rightarrow$MIMIC-CXR preservation. Model colours in (a) match the complete labels in (b).}
\label{fig:shift}
\end{figure}

Under CheXpert$\rightarrow$MIMIC-CXR (\cref{fig:shift}), MedSigLIP/MedGemma
show small relative AUROC gaps (0.040/0.039), and medically pretrained encoders
retain 0.717 macro-AUROC versus 0.540 for generalists (Welch $p=0.0029$).
Preservation is not uniform within the medical group: LLaVA-Med falls to 0.504,
so provenance alone does not secure it. The site shift therefore exposes
preservation differences that clean source-domain AUROC does not reveal.

\subsection{Calibration Repair and Corruption-Induced Miscalibration}
\label{sec:calibration-results}
\label{sec:calib-findings}
\begin{figure}[tbp]
\centering
\includegraphics[width=0.52\linewidth]{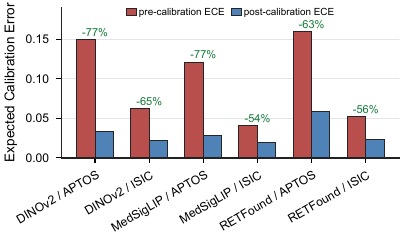}
\caption{\textbf{Post-hoc calibration (E2).} ECE falls by 54--77\% at preserved AUROC.}
\label{fig:calibration}
\end{figure}

Across six primary tests, temperature, vector, isotonic, or Dirichlet scaling
reduces ECE by 54--77\% (mean 0.067, 95\% CI $[0.024,0.110]$) while preserving
AUROC (mean $\Delta$AUROC approximately $+0.001$). DINOv2/APTOS improves
0.150$\rightarrow$0.034 and MedSigLIP/APTOS 0.121$\rightarrow$0.028
(\cref{fig:calibration}). Because calibration changes the probability mapping,
it can repair confidence but cannot recover lost discrimination or transfer.

Corruption exposes a complementary failure mode: class ranking can remain useful while
probability estimates degrade. APTOS LLaVA-Med shows mean ECE increase 0.128
(worst 0.418); on ISIC, MedGemma/MedSigLIP remain strong discriminators while ECE
rises 0.113/0.094, whereas CheXpert is comparatively stable throughout
(maximum 0.038). Discrimination loss and calibration loss are only moderately
associated across the 43 cells ($r=0.63$), so the failure mode is invisible to an
AUROC-only notion of robustness.

\begin{figure}[tbp]
\centering
\begin{subfigure}[b]{0.42\linewidth}
  \centering
  \includegraphics[width=\linewidth]{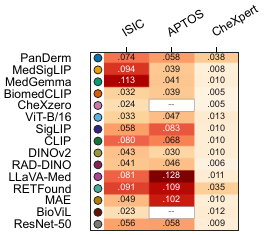}
  \caption{Mean ECE increase.}
\end{subfigure}\hfill
\begin{subfigure}[b]{0.56\linewidth}
  \centering
  \includegraphics[width=\linewidth]{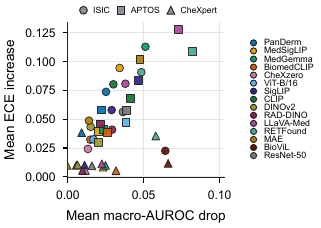}
  \caption{AUROC loss vs. ECE increase.}
\end{subfigure}
\caption{\textbf{Corruption-induced miscalibration (E4).} The losses are moderately associated ($r=0.63$). Colour = encoder, with the complete model key at right; shape = dataset.}
\label{fig:corruption}
\end{figure}

\subsection{Representation Depth and Supervision Efficiency}
\label{sec:representation-results}
\label{sec:repr-findings}
Intermediate features win 5/8 depth sweeps: RAD-DINO gains 1.3--1.6 AUROC points
around 50\% depth, DINOv2 prefers 75\%, and shallower MedSigLIP/MedGemma features
can lower ECE. Final blocks are therefore not uniformly optimal clinical
representations, making layer choice a low-cost but consequential evaluation
decision.

Medical pretraining shows its largest relative advantage in the low-label regime. With 1\% of
labels, MedGemma reaches 0.884 AUROC on APTOS and MedSigLIP/MedGemma/PanDerm
reach 0.819/0.815/0.806 on ISIC. On CheXpert, six medically pretrained encoders
meet ViT-B/16 at 1\%, whereas MAE and ResNet-50 never cross it. Crossover
analysis complements relative-retention LE by expressing supervision demand
against a common reference.

\subsection{Prompting versus Feature Probing}
\label{sec:prompt-results}
\label{sec:prompt-findings}
MedGemma exceeds prompted LLaVA-Med on APTOS/ISIC/CheXpert
(74.7/51.0/82.0\% vs. 49.3/17.8/79.5\%) but remains below frozen-feature
probing; one APTOS chain-of-thought setting parses only 64.6\% of cases.
Representation quality and the generative prediction interface are therefore
distinct sources of reliability variation.

\section{Discussion and Conclusion}
CRS-Bench frames encoder selection as a controlled multi-objective comparison.
Because discrimination is one profile component, positive AUROC--CRS association
is expected; the key result is that adding calibration, label efficiency, and
robustness changes 21 of 105 pairwise choices. Together with paired-seed
uncertainty, this supports a stable leading tier and profile-specific trade-offs
rather than a statistically resolved universal winner.

The reference-relative construction separates score stability from ranking
transfer. Panel perturbations quantify sensitivity to reference composition,
whereas fixed-anchor insertion keeps previously reported scores invariant to new
candidates. LoRA and held-out data instead test whether the frozen ordering
persists after representation or distribution change, delimiting transfer scope
without altering the within-suite selection objective.

Encoder analyses further show why the profile should accompany CRS. Specialty
pretraining improves matched-domain discrimination without uniform cross-domain
gains, while broad medical pretraining is comparatively consistent in this
panel under label scarcity and the observed radiology shift. Calibration is
substantially repairable post hoc, but corruption can preserve AUROC while
degrading probability quality; representation depth and prompting add variation
beyond final-layer clean AUROC.

The scope is classification across three primary domains and one observed
radiology shift; controlled perturbations model acquisition/image quality, and
BioViL/CheXzero lack APTOS frozen cells. CRS is a benchmark-level comparative
summary rather than a clinical-utility endpoint; prospective validation,
subgroup analysis, operating-point selection, and human--AI interaction remain
outside this protocol.
CRS-Bench enables controlled, reference-relative multi-objective selection of
medical image encoders beyond clean-test discrimination. Across 15 encoders,
21 of 105 AUROC--CRS pairwise choices reverse, and paired-seed resampling
supports a stable leading tier rather than a single resolved leader. The
resulting ranking reflects encoder reliability profiles rather than arbitrary
comparison effects: it remains stable under reference-panel perturbations,
while fixed-reference anchoring prevents previously reported scores from
changing as new encoders are added.

CRS-Bench further shows that reliability dimensions capture complementary
properties: specialization, calibration, label efficiency, and robustness
reveal differences that clean AUROC alone cannot characterize. Its scope is a
benchmark-level comparative summary rather than a clinical-utility endpoint,
and future work should extend evaluation to additional modalities and
prospective settings. Within these limits, reliability profiles should
accompany, rather than replace, scalar performance metrics when selecting
medical image encoders.

\begingroup
\setlength{\bibsep}{0.5pt plus 0.5pt}
{\small
\bibliographystyle{ieeenat_fullname}
\bibliography{main}
}
\endgroup

\end{document}